\documentclass[conference]{IEEEtran}
\IEEEoverridecommandlockouts

\usepackage{cite}
\usepackage{amsmath,amssymb,amsfonts}
\usepackage{algorithmic}
\usepackage{graphicx}
\usepackage{textcomp}
\usepackage{xcolor}

\usepackage{amsmath,amssymb,amsfonts}
\usepackage{algorithmic}
\usepackage{graphicx}
\usepackage[inline, shortlabels]{enumitem}
\usepackage{tabularx}
\usepackage{caption}
\usepackage{titlesec}
\usepackage[T2A,T1]{fontenc}
\usepackage[english]{babel}
\usepackage{ragged2e}
\addto\extrasenglish{%
}
\usepackage{pifont}
\usepackage{footmisc}
\usepackage{multirow}

\usepackage{physics}
\usepackage{amsmath}
\usepackage{tikz}
\usepackage{mathdots}
\usepackage{yhmath}
\usepackage{float}
\usepackage{pdfpages}
\usepackage{stfloats}
\usepackage{cancel}
\usepackage{color}
\usepackage{siunitx}
\usepackage{array}
\usepackage{multirow}
\usepackage{amssymb}
\usepackage{gensymb}
\usepackage{tabularx}
\usepackage{extarrows}
\usepackage{booktabs}
\usetikzlibrary{fadings}
\usetikzlibrary{patterns}
\usetikzlibrary{shadows.blur}
\usetikzlibrary{shapes}

\usepackage{pdfpages}
\usepackage{booktabs}
\usepackage{csquotes}
\usepackage{lipsum}  
\usepackage{arydshln}
\usepackage{smartdiagram}
\usepackage[inkscapeformat=png]{svg}
\usepackage{textcomp}
\usepackage{tabularray}\UseTblrLibrary{varwidth}
\usepackage{xcolor}
\def\BibTeX{{\rm B\kern-.05em{\sc i\kern-.025em b}\kern-.08em
    T\kern-.1667em\lower.7ex\hbox{E}\kern-.125emX}}
\usepackage{cite}
\usepackage{amsmath}
\newcommand{\probP}{\text{I\kern-0.15em P}}
\usepackage{etoolbox}
\patchcmd{\thebibliography}{\section*{\refname}}{}{}{}

\usepackage{hyperref}
\titleclass{\subsubsubsection}{straight}[\subsection]

\newcounter{subsubsubsection}[subsubsection]
\renewcommand\thesubsubsubsection{\thesubsubsection.\arabic{subsubsubsection}}

\titleformat{\subsubsubsection}
  {\normalfont\normalsize\bfseries}{\thesubsubsubsection}{1em}{}
\titlespacing*{\subsubsubsection}
{0pt}{3.25ex plus 1ex minus .2ex}{1.5ex plus .2ex}

\makeatletter
\renewcommand\paragraph{\@startsection{paragraph}{5}{\z@}%
  {3.25ex \@plus1ex \@minus.2ex}%
  {-1em}%
  {\normalfont\normalsize\bfseries}}
\renewcommand\subparagraph{\@startsection{subparagraph}{6}{\parindent}%
  {3.25ex \@plus1ex \@minus .2ex}%
  {-1em}%
  {\normalfont\normalsize\bfseries}}
\def\toclevel@subsubsubsection{4}
\def\toclevel@paragraph{5}
\def\toclevel@paragraph{6}
\def\l@subsubsubsection{\@dottedtocline{4}{7em}{4em}}
\def\l@paragraph{\@dottedtocline{5}{10em}{5em}}
\def\l@subparagraph{\@dottedtocline{6}{14em}{6em}}
\makeatother

\makeatletter
\newcommand{\linebreakand}{%
  \end{@IEEEauthorhalign}
  \hfill\mbox{}\par
  \mbox{}\hfill\begin{@IEEEauthorhalign}
}
\makeatother

\setboolean{@twoside}{false}

\def\BibTeX{{\rm B\kern-.05em{\sc i\kern-.025em b}\kern-.08em
    T\kern-.1667em\lower.7ex\hbox{E}\kern-.125emX}}
\begin{document}

\title{Streamlining Resilient Kubernetes Autoscaling with Multi-Agent Systems via an Automated Online Design Framework
}

\author{

    \IEEEauthorblockN{Julien Soulé}
    \IEEEauthorblockA{\textit{Thales Land and Air Systems, BU IAS}}
    \IEEEauthorblockA{\textit{Univ. Grenoble Alpes,} \\
        \textit{Grenoble INP, LCIS, 26000,}\\
        Valence, France \\
        julien.soule@lcis.grenoble-inp.fr}

    \and

    \IEEEauthorblockN{Jean-Paul Jamont\IEEEauthorrefmark{1}, Michel Occello\IEEEauthorrefmark{2}}
    \IEEEauthorblockA{\textit{Univ. Grenoble Alpes,} \\
        \textit{Grenoble INP, LCIS, 26000,}\\
        Valence, France \\
        \{\IEEEauthorrefmark{1}jean-paul.jamont,\IEEEauthorrefmark{2}michel.occello\}@lcis.grenoble-inp.fr
    }



    \and



    \hspace{2cm}
    \IEEEauthorblockN{Louis-Marie Traonouez}
    \IEEEauthorblockA{
        \hspace{2cm}
        \textit{Thales Land and Air Systems, BU IAS} \\
        \hspace{2cm}
        Rennes, France \\
        \hspace{2cm}
        louis-marie.traonouez@thalesgroup.com}

    \and

    \hspace{1.5cm}
    \IEEEauthorblockN{Paul Théron}
    \IEEEauthorblockA{
        \hspace{1.5cm}
        \textit{AICA IWG} \\
        \hspace{1.5cm}
        La Guillermie, France \\
        \hspace{1.5cm}
        paul.theron@orange.fr}
}

\maketitle

\begin{abstract}
    In cloud-native systems, Kubernetes clusters with interdependent services often face challenges to their operational resilience due to poor workload management issues such as resource blocking, bottlenecks, or continuous pod crashes. These vulnerabilities are further amplified in adversarial scenarios, such as Distributed Denial-of-Service attacks (DDoS). Conventional Horizontal Pod Autoscaling (HPA) approaches struggle to address such dynamic conditions, while reinforcement learning-based methods, though more adaptable, typically optimize single goals like latency or resource usage, neglecting broader failure scenarios.
    We propose decomposing the overarching goal of maintaining operational resilience into failure-specific sub-goals delegated to collaborative agents, collectively forming an HPA Multi-Agent System (MAS). We introduce an automated, four-phase online framework for HPA MAS design: 1) modeling a digital twin built from cluster traces; 2) training agents in simulation using roles and missions tailored to failure contexts; 3) analyzing agent behaviors for explainability; and 4) transferring learned policies to the real cluster.
    Experimental results demonstrate that the generated HPA MASs outperform three state-of-the-art HPA systems in sustaining operational resilience under various adversarial conditions in a proposed complex cluster.
\end{abstract}

\begin{IEEEkeywords}
    Adversarial, Horizontal Pod Autoscaling, Multi-Agent Reinforcement Learning, Multi-Agent System Design
\end{IEEEkeywords}
component, formatting, style, styling, insert.

\section{Introduction}
\label{sec:introduction}

Cloud-native critical systems are increasingly reliant on Kubernetes to orchestrate and manage interdependent services~\cite{Pahl2019}. HPA is a widely adopted mechanism to dynamically adjust the number of pods based on resource usage, enabling systems to handle highly dynamic workloads~\cite{Toka2020}. However, failures such as pod crashes, resource contention, and bottlenecks can severely jeopardize the performance of all of the cluster's functionalities we globally refer to as operational resilience~\cite{burns2016borg}. Worse, these failures may be exploited by attackers to degrade performance or induce outages, as seen in adversarial contexts like DDoS attacks~\cite{David2021}.

Although DDoS attacks may seem unlikely in isolated enterprise clusters, many organizations expose services via ingress gateways, making them viable targets. Internally, DDoS-like overloads can also result from misconfigurations, compromised devices, or red-teaming exercises. Accounting for such adversarial scenarios aligns with cybersecurity best practices and helps ensure robust, fault-tolerant autoscaling.

In such adversarial scenarios, malicious actors can exploit scaling mechanisms, exposing the limitations of conventional HPA systems. Modern approaches have sought to address these gaps using Reinforcement Learning (RL), where an agent optimizes a single global goal such as minimizing latency or resource usage~\cite{Gari2021}. While these methods demonstrate adaptability, they often fall short in handling diverse failure scenarios to maintain \textit{Quality of Service} (QoS)~\cite{Liu2024}. For example, prioritizing responses to cascading pod crashes during an attack may be far more critical than reducing latency. These challenges highlight the need for an autoscaling system capable of dynamically balancing multiple sub-goals to maintain all of the QoS to maximize operational resilience.

Achieving this shift from single-goal optimization to a multi-goal approach is complex~\cite{Shoham2009MAS}. In real-world scenarios, operational resilience is not reducible to a single objective: optimizing for low latency may conflict with ensuring high availability or limiting resource overprovisioning. During adversarial conditions, for example, prioritizing DDoS mitigation may require sacrificing latency, while pod crash recovery demands rapid reallocation of resources. A single RL agent struggles to address such priorities due to the difficulty of coordinating responses to diverse failures~\cite{Jennings1998}.

In contrast, MASs offer a promising paradigm by decomposing the overarching operational resilience maximization goal into sub-goals handled by specialized agents~\cite{Shoham2009MAS}. Each agent focuses on a specific failure mode or performance objective, which facilitates specialization, improved coordination, and scalable policy learning under dynamic workloads.

Considering an adversarial scenario, each defender can collaboratively contribute to complementary scaling actions to reach its own sub-goal, enabling more resilient and context-specific responses face to an attacker~\cite{Jennings1998}. We refer to the set of these collaborative as an HPA MAS. An HPA MAS actually builds upon the cyberdefense framework of Autonomous Intelligent Cybersecurity Agents (AICAs), which can be viewed as agents with specialized roles and missions collaboratively defending systems against attackers~\cite{Kott2018}.

However, designing HPA MASs tailored to a cluster presents significant challenges such as the need for detailed cluster knowledge, the time-consuming nature of manual design processes, and the difficulty of ensuring optimal agent behavior. Moreover, cluster changes require repeating the design process, increasing operational costs and complexity.

Among methdological works, we inspired from the \textit{Assisted MAS Organization Engineering Approach} (AOMEA)~\cite{soule2024aomea} that shows to align the most with automation and safety challenges. Based on AOMEA, we propose the \textit{Kubernetes Autoscaling with Resilient Multi-Agent system} (KARMA) to automate the design and implementation process through four sequential phases:
\begin{enumerate*}[label=\textbf{\arabic*)}, itemjoin={;\quad }]
    \item \textbf{Modeling}: Creating a digital twin of the cluster from real-world traces to simulate failure scenarios
    \item \textbf{Training}: Training agents in simulation using roles and missions that integrate explicit rule-based and guidance strategies
    \item \textbf{Analyzing}: Validating trained agents' behaviors and extracting design insights through empirical analysis
    \item \textbf{Transferring}: Running trained agents to apply their learned behaviors via the real Kubernetes API.
\end{enumerate*}

This framework enables to iteratively updates the simulation model with newly collected traces, enabling adaptation to cluster changes. We validated our approach on adversarial scenarios from the "Chained Service" Kubernetes environment. The MASs were generated with minimal manual intervention and demonstrate originality. They compete with state-of-the-art HPA systems, including AWARE~\cite{aware2023}, Gym-HPA~\cite{gymhpa2022}, and Rlad-core~\cite{Rossi2019}, in maximizing operational resilience.

The remainder is structured as follows:
\autoref{sec:related_work} reviews existing HPA techniques and their limitations in dynamic environments.
\autoref{sec:proposed_approach} details our framework leveraging related concepts for each phase.
\autoref{sec:experiments} describes the experimental setup.
\autoref{sec:results} presents and discusses results.
\autoref{sec:conclusion} concludes and provides future directions.


\section{Related Work}
\label{sec:related_work}

\begin{table}[h!]
    \centering
    \caption{A KARMA overview regarding selected HPA Systems}
    \label{tab:autoscaling_criteria}
    {\footnotesize
    \renewcommand{\arraystretch}{1.1}
    \begin{tabular}{>{\raggedright\arraybackslash}m{1.3cm}>{\centering\arraybackslash}m{0.6cm}>{\centering\arraybackslash}m{0.6cm}>{\centering\arraybackslash}m{0.6cm}>{\centering\arraybackslash}m{0.6cm}>{\centering\arraybackslash}m{0.6cm}>{\centering\arraybackslash}m{0.6cm}>{\centering\arraybackslash}m{0.6cm}>{\centering\arraybackslash}m{0.6cm}>{\centering\arraybackslash}m{0.6cm}}
    \hline
    \textbf{Criterion} & \vspace{-0.3cm}\textbf{\cite{gymhpa2022}} & \vspace{-0.3cm}\textbf{\cite{aware2023}} & \vspace{-0.3cm}\textbf{\cite{Rossi2019}} & \vspace{-0.3cm}\textbf{\cite{QoSRL}} & \vspace{-0.3cm}\textbf{\cite{Zhou2024}} & \vspace{-0.3cm}\textbf{\cite{KOSMOS}} & \vspace{-0.3cm}\textbf{\cite{COPA}} \\
    \hline
    \hline
    Adversarial Scenarios & No & Partial & No & No & No & No & Partial \\
    \hline
    Multi-goal & No & Yes & Partial & Yes & No & Yes & No \\
    \hline
    Automation & High & Mid. & Mid. & High & Mid. & Mid. & Mid. \\
    \hline
    Learning & Yes & Yes & Yes & Yes & No & No & No \\
    \hline
    MAS & No & No & No & No & No & No & No \\
    \hline
    Simulation & Yes & No & Yes & Yes & No & No & No \\
    \hline
    Real env. & No & Yes & Yes & Yes & Yes & Yes & Yes \\
    \hline
    Explainable & No & No & No & No & No & No & No \\
    \hline
    Adaptation & High & Mid. & Mid. & Mid. & High & High & Mid. \\
    \hline
    Safety Guarantees & No & No & No & No & No & No & No \\
    \hline
    \end{tabular}%
    }
  \end{table}

\noindent Autoscaling in Kubernetes has traditionally relied on metrics-based approaches, such as the default Kubernetes Horizontal Pod Autoscaler (KHPA), which adjusts the number of pods based on CPU and memory utilization~\cite{Carrion2022}. While effective for basic scaling, such methods fail to address dynamic or adversarial workloads, as they rely on reactive, threshold-based rules~\cite{Tran2022}. To overcome these limitations, recent research has turned to Machine Learning (ML) and RL.

Three RL-based systems stand out for their innovative approaches, applicability, and relevance:
\begin{itemize}
    \item \textbf{Gym-HPA}~\cite{gymhpa2022} serves as a benchmark RL environment, enabling experimentation with various RL algorithms. It excels in adaptability to simulated workloads with a high degree of automation but lacks multi-goal support, explainability, and real-world applicability
    \item \textbf{AWARE}~\cite{aware2023} incorporates RL to optimize autoscaling decisions while balancing QoS goals, such as response time and throughput. It partially considers adversarial scenarios but struggles with high automation levels and multi-agent coordination
    \item \textbf{Rlad-core}~\cite{Rossi2019} focus on both horizontal and vertical scaling introducing a self-adaptive system that dynamically adjusts resource allocation in response to workload variations, optimizing performance while minimizing costs.
\end{itemize}

These systems highlight significant progress in RL-based autoscaling but share common limitations: a lack of comprehensive adversarial adaptability, limited support for MAS, and no explicit focus on explainability or safety guarantees.
%
Other notable systems combine ML or rule-based strategies with traditional autoscaling:
\begin{enumerate*}[label={}, itemjoin={;\quad }]
    \item \textbf{QoS-Aware RL}~\cite{QoSRL} focus on maintaining QoS under dynamic workloads but does not integrate seamlessly with Kubernetes-native features or consider adversarial scenarios
    \item \textbf{AHPA}~\cite{Zhou2024} and \textbf{KOSMOS}~\cite{KOSMOS} explore adaptive and combined vertical-horizontal scaling strategies, offering high adaptability but lacking learning capabilities
    \item \textbf{COPA}~\cite{COPA} emphasize combined metrics-based autoscaling but remains reactive and limited in adversarial scenarios.
\end{enumerate*}

KARMA addresses key gaps in \textbf{(1) operational resilience} for autoscaling by introducing an automated HPA MAS design framework. Unlike conventional approaches, which often fail under \textbf{(2) adversarial conditions}, KARMA decomposes maintaining operational resilience into failure-specific missions and roles, enabling agents to handle coordinate response to \textbf{bottlenecks}, \textbf{resource contention}, \textbf{DDoS}, and \textbf{pod crashes}. It combines \textbf{(3) digital twin modeling} with \textbf{(4) automated MAS generation} via Multi-Agent Reinforcement Learning (MARL) integrating \textbf{constraint satisfaction} to these roles and missions, streamlining HPA MAS design with minimal manual intervention. Leveraging this decomposition, KARMA seeks for a better \textbf{(5) adaptability} while also enabling a better \textbf{(6) explainability} as for decision making in the whole HPA MAS.

\section{KARMA: A framework for HPA MAS design and development}
\label{sec:proposed_approach}

This section overviews the KARMA framework to helps in designing a HPA MAS then details each one of its phase.

\subsection{KARMA Overview}

\begin{figure}[h!]
    \centering
    \input{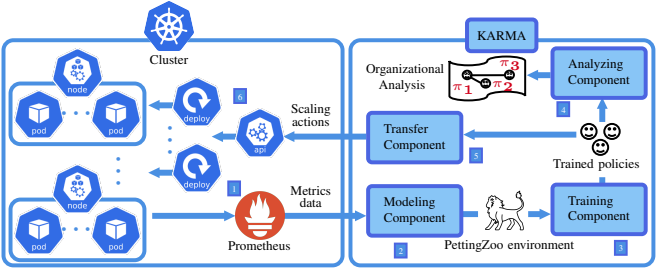}
    \caption{Overview of the KARMA framework in use with a Kubernetes cluster}
    \label{fig:karma_architecture}
\end{figure}

As illustrated in \autoref{fig:karma_architecture}, the KARMA framework operates alongside the Kubernetes cluster, comprising \textbf{worker nodes} that host \textbf{pods} (the atomic unit in Kubernetes containing \textbf{containers} running the actual processes). Pods are organized into \textbf{services} and managed by \textbf{deployments} to update the number of pod refered to as \textbf{replica}. The KARMA framework functions as a separate software layer, interfacing with both the Kubernetes API and Prometheus.

\textbf{1)} Metrics related to availability are gathered as states by \textit{Prometheus}~\cite{prometheus}, a widely adopted time-series metrics database, and processed by KARMA's \textbf{Modeling Component}.

\textbf{2)} Collected states are used to construct a \textit{digital twin} of the cluster as a state transition model. A reward function, defined as a weighted sum of QoS-specific sub-rewards, drives the operational resilience. The digital twin provides a controlled simulation environment to train agents safely without risking disruptions in the real cluster.

\textbf{3)} The \textbf{Training Component} trains agents to maximize rewards to improve operational resilience. Agents are optionally guided by \textit{roles} (constraints shaping their actions) and \textit{missions} (incremental goals facilitating policy convergence), following the AOMEA methodology~\cite{soule2024aomea}.

\textbf{4)} The \textbf{Analyzing Component} visualize the learned policies through trajectory clustering and hierarchical visualization, ensuring interpretability, alignment with goals, and resilience to dynamic workloads.

\textbf{5)} The \textbf{Transfer Component} deploys trained policies to the real Kubernetes cluster via the Kubernetes API, executing replica adjustments \textbf{(6)}. Continous interactions between agents and the cluster enables enriching the digital twin with newly collected traces, eventually updating the agents' policy regularly to meet environment's changes.

KARMA integrates simulation-based learning with real-world Kubernetes operations in a closed-loop process. Metrics collected from the cluster guide policy updates, while trained agents' decisions are applied back to the cluster. This iterative process aims to ensure ongoing adaptability, robustness, and resilience under diverse conditions.

\subsection{Modeling}

In this phase, we assume the initial defender and attacker agents have applied various action in the cluster, leading to a collection of representative amount of traces. Relying on the formalization of the environment as a zero-sum \textbf{Stochastic Game (SG)}~\cite{shapley1953stochastic}, we can provide a near-realistic simulation environment from these collected traces. The SG is characterized by the tuple $\mathcal{SG} = (\mathcal{A}, S, A, T, R, \gamma)$, where $\mathcal{A} = \{\mathcal{A}_d, \mathcal{A}_a\}$ is the agents set comprising $n = |\mathcal{A}_d|$ defender agents and one single attacker agent in $\mathcal{A}_a$; and $\gamma \in [0, 1]$ is the discount factor for future rewards.

\noindent \paragraph{\textbf{State Space}} $S$ is the state space of the Kubernetes cluster. A state denoted as $s \in S$, consists of metrics to characterizing system performance for each of the $d = |D|$ micro-service deployments:
$$
s = (n_{id}, d_{dep}, d_{des}, d_{err}, d_{rem}, r_{cpu}, r_{ram}, t_{in}, t_{out})^d
$$
$n_{id} \in \mathbb{N}$: the deployment number; \quad
$d_{dep} \in \mathbb{N}$: the number of deployed pods; \quad 
$d_{des} \in \mathbb{N}$: the number of desired pods; \quad
$d_{err} \in \mathbb{N}$: the number of failed pods; \quad
$d_{rem} \in \mathbb{N}$: the number of remaining requests to be processed in the queue; \quad
$r_{cpu} \in \mathbb{R}$: the total aggregated CPU (in m) of the pods; \quad
$r_{ram} \in \mathbb{R}$: the total aggregated memory (in Mi) of the pods; \quad
$t_{in} \in \mathbb{R}$: the average received traffic (in Kbps); \quad
$t_{out} \in \mathbb{R}$: the average transmitted traffic (in Kbps).


\noindent \paragraph{\textbf{Action Space}} $A = A_d^n \times A_a$ is the action space with $A_d$ and $A_a$ are the action spaces for a defender and the attacker agents respectively:

\vspace{0.3cm}

\indent\begin{minipage}{0.15\linewidth}
    (1)
\end{minipage}
\begin{minipage}{0.9\linewidth}
    \raggedright
    $\displaystyle a_d \in A_d = (service\_id, replica\_change)$
\end{minipage}

\vspace{0.3cm}

\indent $\mathbf{service\_id} \in \mathbb{N}$ identifies the target service (through deployment), and $\mathbf{replica\_change} \in [-\alpha, +\alpha]$ indicates the change as for pod replica number. Actions from this space are one-hot encoded as a Box Gym Space~\cite{openAIGymActionSpaces}: for example, the defender actions $(2,1)$, $(0,-2), (1,0)$ mean the services with id numbers equal to $2$, $0$, and $1$ have their respective replica numbers changed by adding $1$, $-2$, $0$.

\

\indent\begin{minipage}{0.06\linewidth}
    (2)
\end{minipage}
\begin{minipage}{0.9\linewidth}
    \raggedright
    $\displaystyle a_a \in A_a = (\text{entry\_point\_id}, \text{rate\_change}, \text{data\_change})$
\end{minipage}

\vspace{0.3cm}

\indent $\mathbf{entry\_point\_id} \in \mathbb{N}$ specifies the service entry point;
$\mathbf{rate\_change} \in \{\textit{high\_decrease}, \textit{low\_decrease}, \textit{no\_change}, \allowbreak \textit{low\_increase}, \allowbreak \textit{high\_increase}\}$ changes the incoming traffic based on a factor $\kappa$; and $\mathbf{data\_change} \in \{\textit{no\_alteration}, \allowbreak \textit{low\_alteration}, \allowbreak \textit{high\_alteration}\}$ specifies the degree of data alteration based on factor $\sigma \in [2,\infty[$. Actions from this space are one-hot encoded as a Box Gym space: for example, the attacker actions $(0,1,2), (2,-1,0), (1,2,1)$ mean that entrypoint services id number 0, 2, and 1 would have their respective traffic-in rates increased by $1 \times \kappa, -1 \times \kappa, 2 \times \kappa$, and respective probabilities to crash due to data alteration are changed by $\frac{2}{\sigma}, \frac{0}{\sigma}, \frac{1}{\sigma}$.

\noindent \paragraph{\textbf{Reward Functions}} $R = \{R_d, R_a\}$, with $R_d: S \times A \to \mathbb{R}$ and $R_a: S \times A \to \mathbb{R} = - R_d$ are respectively the reward function for the defender agents based on operational resilience and the attacker one.
To measure operational resilience, we use the linear combination of the following metrics:
\begin{itemize}
    \vspace{0.15cm}
    \item $\text{Success Rate } (sr) : \frac{\text{Successful Requests}}{\text{Total Received Requests}}$
    \vspace{0.15cm}
    \item $\text{Pod Failure Rate } (pfr) : \frac{\text{Failed Pods}}{\text{Total Deployed Pods}}$
    \vspace{0.15cm}
    \item $\text{Latency Ratio } (lr) : \min\left(1,\frac{\text{Measured Latency}}{\text{Maximum Acceptable Latency}}\right)$
    \vspace{0.15cm}    
    \item $\text{Entry Point Availability } (epa) : \frac{\text{Available Entry Points}}{\text{Total Entry Points}}$
    \vspace{0.15cm}
    \item $\text{Traffic Capacity Ratio } (tcr) : \min\left(1, \frac{\text{Outgoing Traffic}}{\text{Expected Traffic}}\right)$
\end{itemize}

\vspace{0.3cm}

$\text{Operational Resilience }: or(s) = w_1 \times sr
\allowbreak + w_2 \times (1 - pfr)
\allowbreak + w_3 \times (1 - lr)
\allowbreak + w_4 \times epa
\allowbreak + w_5 \times tcr
\text{ where } (w_1, w_2, w_3, w_4, w_5) \text{ are relative weights.}$

\

Then, the reward function is formalized as follow:

$$
\begin{cases} 
    R_a(s, a_d, a_a) = -R_d(s, a_d, a_a) & \\
    R_a(s, a_d, a_a) = or(s)
\end{cases}
$$

where $s$ is the current state after applying actions, and the weights $(w_1, w_2, w_3, w_4, w_5)$ are empirically tuned to favor a balanced global functioning in the cluster.

\noindent \paragraph{\textbf{Transition Modeling}} $T: S \times A \rightarrow S$ is the real state transition function dictating the next state when the joint actions of the defender agents and the attacker's one are applied. Relying a representative set of collected transitions $\mathcal{T} = \langle(s, a_d^n, a_a, s')_{t\in \mathbb{N}}\rangle$ over a time window, we can form a partial state transition function $\hat{T}_t$ defined as:
$$
\hat{T}_t(s, a_d^n, a_a) =
\begin{cases} 
    s' & \text{if } (s, a_d^n, a_a, s') \in \mathcal{T} \\
    \emptyset & \text{ otherwise}
\end{cases}
$$

To cover no recorded transition, we introduce a Multi-Layer Perceptron (MLP)-based approximator $\hat{T}_a$ to learn from collected transitions and predict the next likely state. The choice of an MLP is motivated by its universal approximation capabilities and the assumption that the next state of a Kubernetes cluster depends only on the current state and the chosen actions. This assumption holds because:
\begin{enumerate*}[label={\roman*)}, itemjoin={;\quad }]
    \item Kubernetes autoscaling decisions are primarily dictated by real-time resource metrics (CPU, memory, network) which do not depend on historical states beyond a small time window
    \item Previous works in reinforcement learning for autoscaling~\cite{Gari2021} have demonstrated that Markovian models effectively approximate real-world cluster behavior.
\end{enumerate*}
The MLP approximator has three hidden layers, striking a balance between expressiveness and computational efficiency. The input and output layer dimensions correspond to the size of the state space, while each hidden layer consists of 128 neurons. Rectified Linear Unit (ReLU) activation functions are applied to the hidden layers, and a linear activation function is used at the output layer to produce the predicted next state. The model is optimized using the Adam optimizer with a learning rate of $10^{-3}$, minimizing the Mean Squared Error loss function, expressed as
$$
\mathcal{L} = \frac{1}{N} \sum_{i=1}^N |T(s_i, a_{d,i}, a_{a,i}) - s'_i|^2
$$
where $N$ is a number of representative transitions to ensure generalization.

The complete modeled transition function $\hat{T}$ is defined as:
$$
\hat{T}(s, a_d, a_a) = 
\begin{cases} 
\hat{T}_t(s, a_d, a_a) & \text{if } (s, a_d, a_a) \in \text{Domain}(\hat{T}_t), \\
\hat{T}_a(s, a_d, a_a) & \text{otherwise}.
\end{cases}
$$

\noindent \paragraph{\textbf{Digital Twin Environment}} The defined action and state spaces with the approximated transition function $\hat{T}$, combined with the reward functions $R_d$ and $R_a$, forms the basis of the digital twin environment implemented using the PettingZoo library~\cite{Terry2021}. This environment enables simulating the cluster from the defined SG, offering a safe space for defender agents to explore various strategies against the attacker agent.

\subsection{Training}
\label{sec:training}

In this phase, MARL algorithms are applied within the modeled environment to enable agents to learn by maximizing cumulative rewards. As suggested in AOMEA~\cite{soule2024aomea}, we leverage the $\mathcal{M}OISE^+$ organizational model to bring ways to control/guide the MARL training. In the KARMA framework it results in a decomposition of the overarching goal of \textit{operational resilience} into sub-goals. Each sub-goal is assigned to a specific agent as a mission, while roles define rule-based strategies to guide agent operations.\\

\noindent \textbf{Agent Roles and Missions}

\

A \textbf{role} is formally represented by a \textbf{Role Action Guide} (RAG), which restricts an agent's permissible actions:
$$
rag(h, \omega) = (\{a_1, a_2, \dots, a_i\}, ch)
$$
where $h$ represents the trajectory or history, \(\omega\) is the agent's observation, and \(ch \in \{0,1\}\) is the constraint hardness. A hard constraint (\(ch = 1\)) strictly limits the agent's available actions to authorized ones, while a soft constraint (\(ch = 0\)) allows for exploratory actions but adjusts rewards with bonuses or penalties based on compliance with the authorized action set. For example, the \textit{Bottleneck Manager} has a role restricting its actions to modifying pod replicas within a specific service graph, ensuring it does not affect unrelated services. If a bottleneck is detected in Service A that causes delays in Service B, the agent adheres to its role by increasing the pod replicas of Service A to resolve the issue without overstepping constraints.

\

A \textbf{mission} is a set of intermediate goals designed to assist in achieving the overarching operational resilience goal. A goal is represented by a \textbf{Goal Reward Guide} (GRG), which incentivizes achieving the expected outcome for this goal:
$$
grg(h) = r_b,
$$
where \(h \in H\) represents the current agent trajectory and \(r_b \in \mathbb{R}\) is the associated reward bonus or penalty. This mechanism narrows the optimization focus to critical resilience tasks. For instance, the \textit{DDoS Manager} is assigned a mission to maintain service availability during DDoS attacks. It earns a reward bonus for ensuring that a predefined percentage of incoming requests is successfully handled, despite the increased load, by dynamically adjusting replicas. The mission aligns the agent's focus on balancing load while avoiding resource over-provisioning.\\

By defining specific \textbf{(role, mission)} pairs, the KARMA framework assigns specialized tasks to agents to tackle distinct challenges in chained Kubernetes services. These roles and missions enable distributed yet coordinated policy learning, ensuring that agents align their actions to maximize overall \textit{operational resilience}. For example, while the \textit{Bottleneck Manager} alleviates bottlenecks to ensure steady service throughput, its actions complement those of the \textit{DDoS Manager}, which adjusts replicas to mitigate traffic surges. This interdependence highlights the importance of coordinating roles and missions to address shared challenges.

\paragraph*{Algorithms and training pipeline}

KARMA integrates \textit{Multi-Agent Proximal Policy Optimization}~\cite{Yu2022} (MAPPO), an Actor-Critic MARL algorithm where the centralized critic provides global feedback, helping agents learn cooperative strategies by optimizing a shared value function. Meanwhile, decentralized actors ensure independent decision-making, preserving realistic execution. This structure fosters emergent collaboration as agents implicitly coordinate through the centralized learning process~\cite{Yu2022}, which is critical in KARMA due to the interdependencies between roles and missions.

The training pipeline in KARMA follows a systematic sequence to optimize agent behaviors. Initially, policies (\(\pi_i\)), roles (RAG), and missions (GRG) are defined for all agents to establish a structured framework. Agents participate in simulation runs, generating trajectories within the digital twin environment. At each step, roles with hard constraints restrict the available actions to authorized ones, while soft constraints influence rewards by applying bonuses or penalties based on compliance. Missions add further reward shaping by incentivizing goal-aligned trajectories. The \textit{Optuna}~\cite{akiba2019optuna} framework is used for Hyper-Parameter Optimization, iteratively refining policies by adjusting hyperparameters~\footnote{} and role specifications to improve convergence and enhance system resilience.

Convergence is deemed successful when the standard deviation of cumulative rewards across episodes falls below a predefined threshold $\mu \in \mathbb{R}$, and cumulative rewards exceed an empirically determined minimum value $\lambda \in \mathbb{R}$.

\subsection{Analysis}
\label{sec:analysis}

This phase aims to interpret agent behaviors as understandable roles and goals, particularly in dynamic, unstructured environments where loosely defined roles and goals lead to broad policy convergence. More complexity arises when agents must coordinate, as in DDoS scenarios, where prioritization emerges implicitly. Understanding these interactions can help ensuring structured coordination and optimal policy convergence~\cite{Shoham2009MAS}.

We propose a manual post-training method to analyze the trajectories of trained agents (comparable to sequences of actions) using unsupervised learning techniques to identify roles, missions, and inter-agent interactions based on a general definition of each, derived from their trajectories.

We also envisioned this method to reduce reliance on domain-specific knowledge by framing role and goal design as an optimization problem, where agent policies evolve within a policy space that is iteratively refined. Training begins with minimal constraints, allowing efficient behaviors to emerge, revealing abstract roles and goals that progressively narrow the search space. With each iteration, newly identified roles and goals further restrict the space, improving precision. Through repeated training, analysis, and refinement, this process aims to design relevant, domain-agnostic roles and goals with minimal manual intervention, enhancing KARMA's generalizability.

\paragraph*{\textbf{Identifying Roles}}

We propose that agents sharing the same role should exhibit similar trajectories. To identify recurring action sequences that define a common role among agents, we apply hierarchical clustering. The distance between these sequences is computed using Dynamic Time Warping~\cite{berndt1994using} (DTW), allowing for variations in timing across test episodes:
\[
d(\tau_i, \tau_j) = \min_{\pi} \sum_{k=1}^{|\pi|} \|a_{t_k}^i - a_{t_k}^j\|_2,
\]
where $\pi$ is the optimal alignment path. Clustering hyperparameters, such as the number of clusters, are empirically tuned to minimize noise and avoid generating rough roles. Clusters are annotated to define abstract roles if they do not match existing ones, such as \textit{Bottleneck Manager} or \textit{DDoS Manager}.

\paragraph*{\textbf{Identifying Missions}}

We propose that agents sharing the same goal should transition through at least one similar state, though possibly via different paths. To identify such similar states, we leverage K-means clustering to group trajectories of performing agents based on the similarity of visited states. Similar states in a cluster are used to define intermediate goals:
\[
g_i = \mathcal{S}_j, \quad \text{where } \mathcal{S}_j = \{s \in \tau_i | \mathbb{P}(s) > \epsilon\}.
\]
Here, $\mathbb{P}(s)$ represents the probability of visiting a state $s$ within successful trajectories. $\epsilon$ represents a threshold used to filter states $s$ based on their probability $\mathbb{P}(s)$ of being visited to minimize noise. Hyperparameters of K-means are empirically optimized to minimize noise. Sampled states are annotated to define abstract goals if they do not match existing ones.

\paragraph*{\textbf{Identifying Inter-Agent Interactions}}

Explainability in KARMA also requires an understanding of how agents coordinate their actions in different contexts. This is achieved by analyzing inter-agent relationships using the concepts of $\mathcal{M}OISE^+$~\cite{hubner2002moise}, which formalizes relations such as:
\begin{enumerate*}[label=\textbf{\arabic*)}, itemjoin={;\quad }]
    \item \textbf{Acquaintance Relations}: Derived from shared observations or reward signals during training, these relations represent which agents are aware of each other's actions.
    \item \textbf{Authority Relations}: Captured through dependency patterns, authority relations prioritize one agent's decisions over another's in specific scenarios. For instance, an specific agent may hold authority during a DDoS attack.
    \item \textbf{Communication Relations}: Inferred from synchronized actions or information-sharing events. For example, agents sharing traffic load data to coordinate replica adjustments during surges.
\end{enumerate*}

\begin{figure}[h!]
    \centering
    \includegraphics[trim=0cm 2cm 0cm 2cm, clip, width=0.5\textwidth]{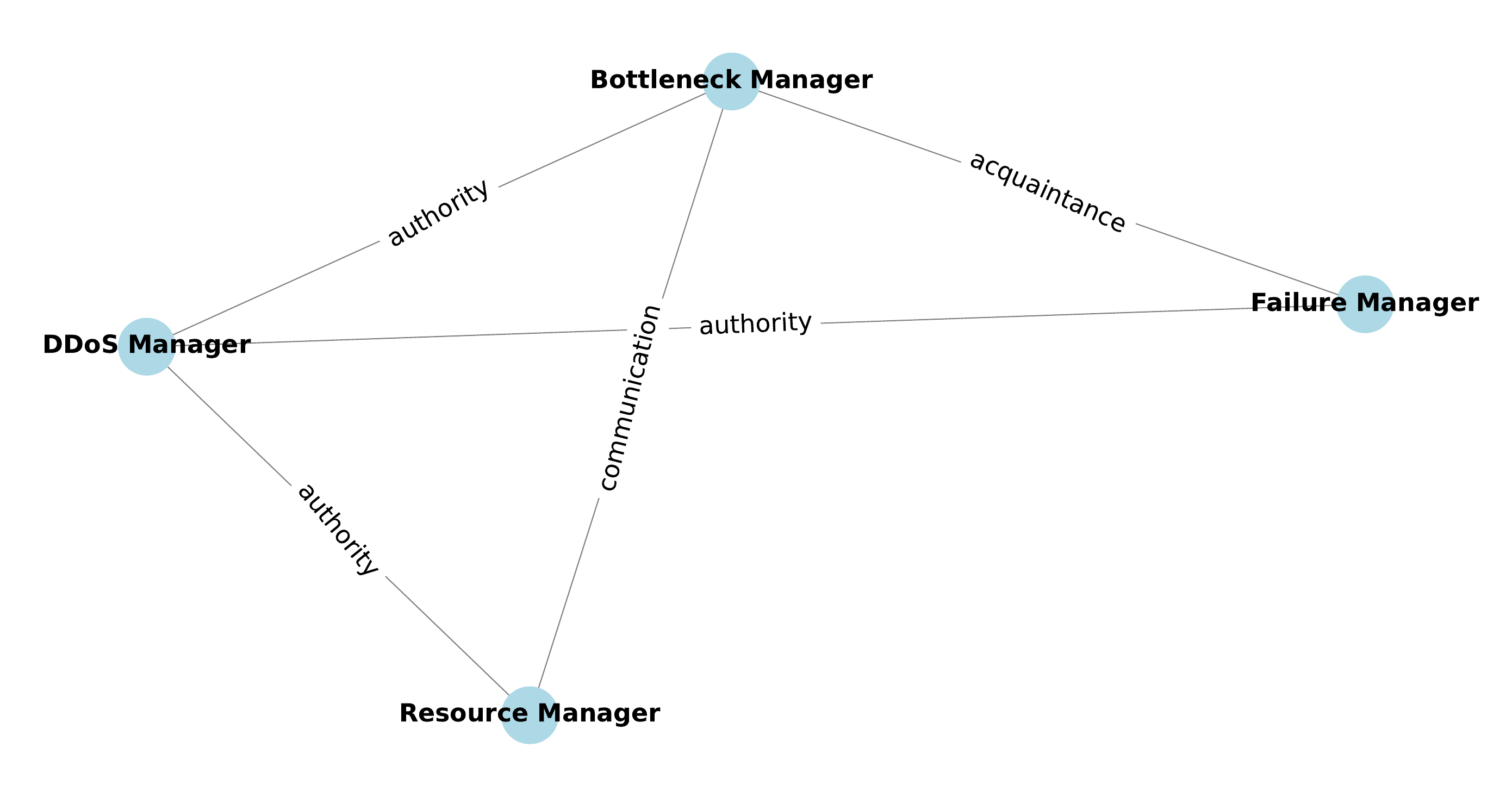}
    \caption{Example of inter-agent graph inferred from trajectories during a DDoS attack: The \textit{DDoS Manager} holds authority over the other agents, coordinating their responses. Additional relations capture observed communication (e.g., Bottleneck and Resource Managers) and awareness links (e.g., acquaintance between Failure and Bottleneck Managers).}
    \label{fig:example_interaction_graph}
\end{figure}

To visualize these relations, a directed graph can be constructed where nodes represent agents and edges represent relations, annotated with interaction type on a time window. An illustrative example is provided in \autoref{fig:example_interaction_graph}.

\subsection{Transfer}
\label{sec:transfer}


In this phase, policies interact with the real cluster:
\begin{enumerate*}[label=\textbf{\arabic*)}, itemjoin={;\quad }]
    \item \textbf{State Collection:} Real-time metrics such as CPU usage, memory consumption, pod status, and network traffic are collected from the Prometheus server~\cite{prometheus}
    \item \textbf{Policy Execution:} Each agent's policy $\pi_i$ computes an action $a_t^i$ based on the current state $s_t$, selecting adjustments such as scaling pod replicas for a deployment
    \item \textbf{Action Application:} The computed actions are sent as API requests, directly modifying deployment.
\end{enumerate*}

Agents in the transfer component continuously interact with the Kubernetes API, generating states that are stored in the modeling component's database. Initially, a large time window is used to collect representative traces for creating a near-realistic digital twin of the Kubernetes cluster. Subsequently, at shorter regular intervals, the modeling component updates the digital twin using the latest trace data. This iterative process aims to ensure agents dynamically adapt to workload.

\noindent \textit{Safety Mechanisms:} To ensure safe policy deployment, KARMA applies several runtime safeguards. Action magnitudes are capped to avoid extreme scaling behaviors, and fallback mechanisms enable reverting to standard KHPA policies if unexpected states are detected. Additionally, policies can be first tested in canary deployments, isolating their effect before full rollout. Monitoring components continuously evaluate agent decisions and trigger alerts if anomalies occur.

%



\section{Experimental Setup}
\label{sec:experiments}

This section outlines the experimental setup for evaluating KARMA's ability to address initially defined gaps.

\subsection{Description of the Kubernetes Cluster and Configuration}

\begin{figure}[h!]
    \centering
    \hspace{-0.4cm}
    \includegraphics[trim=1.8cm 3.3cm 1.25cm 3.5cm, clip, width=0.5\textwidth]{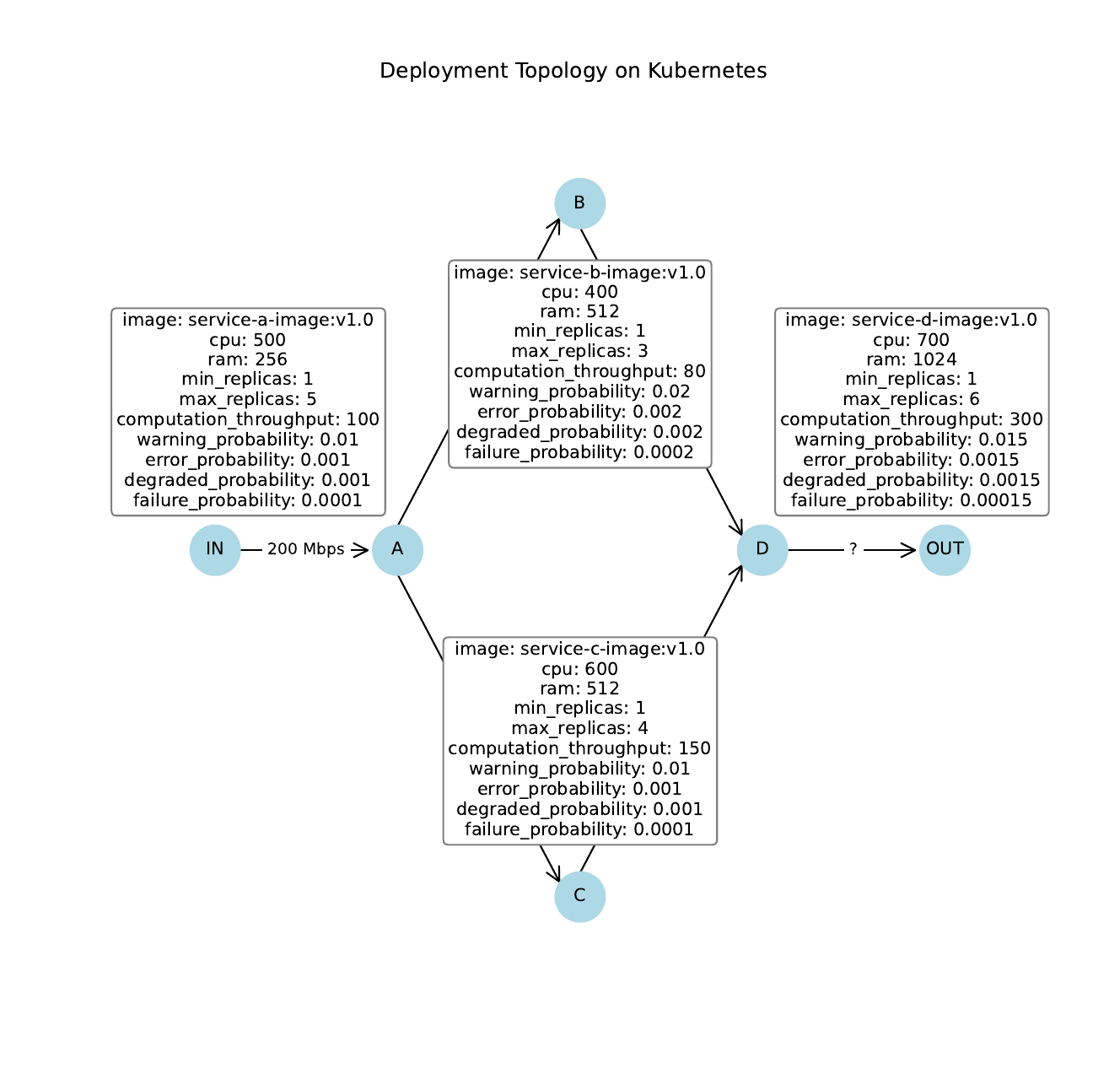}
    \caption{A graph representation of a "Chained Services" cluster with four services}
    \label{fig:chained_services_graph}
\end{figure}

The evaluation environment consists of a Kubernetes cluster simulating a \textbf{Chained Services} (CS) architecture. Each service comprises a set of microservices hosted in pods and managed by deployments. For instance, \autoref{fig:chained_services_graph} illustrates the graph representation of a four services CS cluster. We considered using a cluster characterized by the following specifications:

\begin{itemize}
    \item \textbf{Topology:} Four interconnected running services configured to emulate real-world conditions, including resource contention, bottlenecks, and adversarial scenarios;
    \item \textbf{Failure Simulation:} Bottlenecks and cascading failures are induced by resource-intensive workloads, while adversarial conditions (e.g., DDoS attacks) are emulated using Locust~\cite{locust2021} and random-based custom scripts;
    \item \textbf{Worker Nodes:} 1 worker node with 8 vCPUs, 32 GB RAM, and 1 Gbps network bandwidth. This configuration is suitable for testing purposes on medium-sized clusters;
    \item \textbf{Training Node:} 1 high computing cluster comprising nodes with NVIDIA Tesla V100 GPUs (16GB), Intel Xeon Platinum CPUs (2.3 GHz, 16 cores), 128 GB RAM.
\end{itemize}

\subsection{Implementation of KARMA with CybMASDE}


\footnotetext[2]{\label{lnk:footnote_2}In our implementation by default $\alpha = 3$, $\sigma = 10$, $\kappa = 1$, $ch = 1$, reward weights $(w_1, w_2, w_3, w_4, w_5) = (0.2, 0.2, 0.2, 0.2, 0.2)$, $Q_{\text{threshold}} = 30$ and $U_{\text{threshold}} = 90\%$.
KARMA's source code and other hyperparameters can be found in \url{https://github.com/julien6/KARMA}}

The KARMA framework leverages \textit{Cyber Multi-Agent System Development Environment}~\textsuperscript{\ref{lnk:footnote_2}} (CybMASDE) which is a general assisted-design MAS framework that seamlessly integrates into the KARMA framework.
The framework includes:
\begin{enumerate*}[label=\textbf{\arabic*)}, itemjoin={;\quad }]
    \item \textbf{Digital Twin Modeling:} A simulation environment replicates Kubernetes cluster using real-world traces
    \item \textbf{MARL Training:} \textit{MAPPO}~\cite{Yu2022} is used to train agents in the digital twin environment
    \item \textbf{Organizational Specifications:} Roles and missions defined for agents guide the training process, ensuring coordinated behavior and explainability
    \item \textbf{Deployment Integration:} Trained policies interact with the Kubernetes API to adjust pod replicas in real time.
\end{enumerate*}

\subsection{Roles and Missions for Operational Resilience}

\noindent Following to the $\mathcal{M}OISE^+$~\cite{hubner2002moise} and the AICA architectural insights~\cite{kott2018autonomous}, we implemented four roles to address a specific degradation factor in a QoS.
Each role is associated with a mission, containing a single sub-goal based on metrics.

\noindent \paragraph{\textbf{Bottleneck Manager}} 
The \textit{Bottleneck Manager} role is to monitor services for bottlenecks caused by imbalanced traffic flows. It is based on rules following these metrics:
\begin{enumerate*}[label={}, itemjoin={;\quad }]
    \item \( T_{\text{in}}^i \): Incoming traffic for service \( i \) (Kbps)
    \item \( T_{\text{out}}^i \): Outgoing traffic for service \( i \)
    \item \( Q_{\text{pending}}^i \): Pending requests for service \( i \).
\end{enumerate*}
A bottleneck is detected if: $Q_{\text{pending}}^i > Q_{\text{threshold}} \quad \text{or} \quad T_{\text{in}}^i > \alpha \cdot T_{\text{out}}^i$
where \( Q_{\text{threshold}} \) is the critical queue threshold, and \( \alpha > 1 \) is an amplification factor.

The associated mission aims to minimize the pending queue size to eliminate bottlenecks. The reward function is defined as: $R_{\text{bottleneck}} = - \sum_{i} Q_{\text{pending}}^i$
Agents are rewarded for reducing pending requests, optimizing the throughput~\cite{burns2016borg}.

\noindent \paragraph{\textbf{DDoS Manager}}

The \textit{DDoS Manager} role is to identify DDoS attacks by analyzing traffic anomalies:
\begin{enumerate*}[label={}, itemjoin={;\quad }]
    \item \( R_{\text{rate}} \): Incoming request rate for the cluster.
    \item \( L_{\text{avg}} \): Average observed latency.
    \item \( \Delta T \): Change in traffic volume over a time window \( t \).
\end{enumerate*}
A DDoS attack is detected when:
$R_{\text{rate}} > R_{\text{threshold}} \quad \text{and} \quad \Delta T > \Delta T_{\text{threshold}}$
where \( R_{\text{threshold}} \) is a critical traffic threshold.

The associated mission is to isolate affected services to minimize downtime with this reward function:
$R_{\text{ddos}} = - \left( \text{DownTime} \cdot w_{\text{d}} + L_{\text{avg}} \cdot w_{\text{l}} \right)$
where \( w_{\text{d}} \) and \( w_{\text{l}} \) are weights for downtime and latency, respectively~\cite{Liu2018}.

\noindent \paragraph{\textbf{Failure Manager}}

The \textit{Failure Manager} role is to monitor pod health and eliminates failed pods following this rule:
\begin{enumerate*}[label={}, itemjoin={;\quad }]
    \item \( F_{\text{fail}}^i \): Number of failures for pod \( i \)
    \item \( S_{\text{status}}^i \): Status of pod \( i \) (e.g., \textit{CrashLoopBackOff}).
\end{enumerate*}
A pod failure is detected if:
$F_{\text{fail}}^i > F_{\text{threshold}}$
where \( F_{\text{threshold}} \) is the maximum number of tolerated failures.

The associated mission minimizes downtime caused by repeated failures with this reward function:
$R_{\text{failure}} = - \sum_{i} T_{\text{downtime}}^i$
Agents are incentivized to quickly eliminate and restart failed services.

\noindent \paragraph{\textbf{Resource Manager}}

The \textit{Resource Manager} role is to prioritize critical services when resource contention occurs. The rules are based on:
\begin{enumerate*}[label={}, itemjoin={;\quad }]
    \item \( U_{\text{cpu}}^i \): CPU utilization of service \( i \)
    \item \( U_{\text{mem}}^i \): Memory utilization of service \( i \)
    \item \( P_{\text{priority}}^i \): Priority level of service \( i \) (critical, normal, low).
\end{enumerate*}
Contention is detected if total CPU usage exceeds a threshold:
$U_{\text{cpu}}^{\text{total}} > U_{\text{threshold}}$
Non-critical services are scaled down to free resources:
$\text{Replicas}_{\text{new}}^i = \max\left( \text{Replicas}_{\text{current}}^i - \delta, 1 \right)$

The associated mission ensures critical services by balancing resource usage with this reward function:
$R_{\text{resource}} = - \sum_{i \in \text{Critical}} \left( U_{\text{cpu}}^i + U_{\text{mem}}^i \right)$
Agents are rewarded for prioritizing services while maintaining efficient resource usage~\cite{shahrad2020resource}.

\

\subsection{Experimental Protocol}

\noindent To evaluate KARMA's performance in addressing the six gaps, we propose comparing baselines accross scenarios.

\paragraph{\textbf{Real-Cluster Integration \& Evaluation}}

KARMA couples simulation with real-cluster interaction by building a digital twin from real Kubernetes traces. Trained policies are deployed via the Kubernetes API, influencing the real cluster, while new traces refine the simulation. This ensures real-world applicability with agent training in a safe environment.

\paragraph{\textbf{Experimental Scenarios}}

\noindent Five experimental scenarios are defined to simulate key factors impacting operational resilience in Kubernetes:
\begin{enumerate*}[label=\textbf{\arabic*)}, itemjoin={;\quad }]
    \item \textbf{Bottleneck Resolution:} Simulates scenarios where upstream services overload downstream services to maximize throughput by dynamically scaling replicas
    \item \textbf{DDoS Attack:} Models a sudden surge in traffic aimed at disrupting critical services to detect the attack, isolate affected services, and minimize downtime~\cite{Liu2018}
    \item \textbf{Pod Failures:} Pod crashes are triggered to evaluate the system's ability to restore affected services~\cite{burns2016borg}
    \item \textbf{Resource Contention:} Simulates high resource demand, requiring dynamic prioritization of critical services to maintain overall cluster functionality~\cite{Vhatkar2022}
    \item \textbf{Mixed Scenario:} Combines all scenarios to evaluate the system's adaptability and resilience.
\end{enumerate*}

\paragraph{\textbf{Baselines from the literature}}
\noindent We selected three HPA systems as baselines:
\begin{enumerate*}[label=\textbf{\arabic*)}, itemjoin={;\quad }]
    \item \textbf{AWARE:} An RL-based system that balances response time and throughput~\cite{aware2023}
    \item \textbf{Gym-HPA:} An RL environment for experimentation with various RL algorithms in simulation~\cite{gymhpa2022}
    \item \textbf{Rlad-core}~\cite{Rossi2019} A RL-based simulator which uses machine learning techniques to scale services, most notably Q-learning and Model-based algorithms.
\end{enumerate*}

These baselines have been tested under the same five scenarios using source code when available.

\paragraph{\textbf{Baselines as ablation studies}}

\noindent To isolate the contributions of KARMA's components, ablation studies have been performed following these configurations:
\begin{itemize}
    \item \textbf{With/without MLP:} Evaluates the impact of using an MLP-based transitioner for digital-twin modeling.
    \item \textbf{With/without organizational specifications:} Tests hard and soft organizational constraints during training:
        \begin{itemize}
            \item \textit{Hard constraints:} Strictly enforce roles and missions.
            \item \textit{Soft constraints:} Allow exploratory actions with rewards based on organizational specifications.
        \end{itemize}
    \item \textbf{Multi-agent vs Mono-agent:} Compares a multi-agent configuration with a mono-agent baseline.
\end{itemize}

\paragraph{\textbf{Performance Metrics}}

\noindent For each scenario and baseline, the following metrics are collected:
\begin{enumerate*}[label=\textbf{\arabic*)}, itemjoin={;\quad }]
    \item \textbf{Operational Resilience:} Based on the global reward from the success rate (\%), ratio of pending request (\%), average latency (ms)
    \item \textbf{Adversarial Robustness:} Based on the standard deviation of the reward and the recovery time after DDoS (s), percentage of services remaining available (\%)
    \item \textbf{Digital Twin Accuracy:} Based on the modeled transition model accuracy (\%), computed as the ratio of real cluster performance over the simulation's one
    \item \textbf{Automated MAS Generation:} Based on training convergence time (number of episodes)
    \item \textbf{Adaptability:} Based on the reward standard deviation variance over training episodes on all scenarios (\%)
    \item \textbf{Explainability:} Based on alignment of behaviors with roles/missions when given (\%), and qualitative evaluation of clustering of trajectories.
\end{enumerate*}

\section{Results and Discussion}
\label{sec:results}

This section analyzes the performance of KARMA in addressing the six identified gaps.

\subsection{Gap 1: Operational Resilience}
Operational resilience evaluates the ability of the system to handle failures and maintain high QoS.
\begin{table}[h]
    \centering
    \caption{Operational resilience metrics across all scenarios.}
    \label{tab:operational_resilience}{\footnotesize
    \begin{tabular}{>{\raggedright\arraybackslash}m{2.7cm}>{\centering\arraybackslash}m{1.5cm}>{\centering\arraybackslash}m{1.5cm}>{\centering\arraybackslash}m{1.5cm}}
        \hline
        \textbf{Baseline} & \textbf{Success Rate (\%)} & \textbf{Latency Compliance (\%)} & \textbf{Pending Requests (\%)} \\
        \hline
        KHPA & 64.8 & 58.1 & 20.7 \\
        Gym-HPA & 73.1 & 65.7 & 20.8 \\
        Rlad-core & 77.4 & 70.1 & 15.9 \\
        AWARE & 80.6 & 73.8 & 13.3 \\
        Single-Agent w/o Org. Spec. & 72.6 & 65.4 & 17.0 \\
        Single-Agent w/ Hard Org. Spec. & 80.8 & 72.5 & 15.4 \\
        Multi-Agent w/o Org. Spec. & 87.7 & 81.5 & 9.3 \\
        Multi-Agent w/ Soft Org. Spec. & 82.0 & 74.7 & 15.0 \\
        \textbf{Multi-Agent w/ Hard Org. Spec. (KARMA)} & \textbf{90.9} & \textbf{85.7} & \textbf{5.9} \\
        \hline
    \end{tabular}}
\end{table}
\autoref{tab:operational_resilience} presents a comparison of KARMA against existing baselines. The results show that KARMA achieves the highest success rate (\textbf{90.9\%}), surpassing all baselines including AWARE (80.6\%) and Rlad-core (\textbf{77.4\%}). Similarly, the latency compliance of KARMA is the highest at \textbf{85.7\%}, while the lowest pending requests ratio (\textbf{5.9\%}) suggests efficient handling of workload variations.

\noindent \textit{Statistical Note:} All reported values represent the mean over 10 independent evaluation runs. Although not shown in the tables for brevity, the standard deviation across runs remained consistently low ($\pm$1.8\% for success rate and $\pm$2.1 ms for latency in the mixed scenario), indicating robust outcomes.

KARMA's performance stems from its structured multi-agent coordination, which optimally distributes resources based on failure contexts, avoiding redundant or conflicting scaling actions—issues common in single-agent RL-based autoscalers.
Reactive, threshold-based autoscalers like KHPA and Gym-HPA struggle with dynamic workloads, leading to higher pending requests and lower success rates. AWARE and Rlad-core improve response time and throughput but lack multi-agent coordination, resulting in slower reactions in adversarial scenarios. Single-Agent w/o Organizational Specifications suffers from inefficient resource allocation, while Single-Agent w/ Hard Organizational Specifications benefits from structured decision-making but still lacks distributed coordination.

KARMA's role-based coordination minimizes inefficiencies and enhances decision stability. Its hierarchical decomposition of objectives enables independent yet complementary decisions, leading to more resilient autoscaling.

The results also underscore the value of organizational constraints. Multi-agent systems with soft constraints (82.0\% success rate) outperform single-agent approaches, but KARMA's hard constraints achieve the best results, eliminating conflicting agent behaviors and optimizing scaling.

\subsection{Gap 2: Adversarial Conditions}

Adversarial conditions evaluate the system's robustness against disruptive scenarios such as DDoS attacks.
\begin{table}[h]
    \centering
    \caption{Performance under DDoS scenario.}
    \label{tab:adversarial_conditions}{
        \footnotesize
    \begin{tabular}{>{\raggedright\arraybackslash}m{3.6cm}>{\centering\arraybackslash}m{1.8cm}>{\centering\arraybackslash}m{2cm}}
        \hline
        \textbf{Baseline} & \textbf{Recovery Time (s)} & \textbf{Service Availability (\%)} \\
        \hline
        KHPA & 80.7 & 65.6 \\
        Gym-HPA & 66.2 & 72.6 \\
        Rlad-core & 37.4 & 78.3 \\
        AWARE & 49.5 & 83.6 \\
        Single-Agent w/o Org. Spec. & 60.3 & 72.4 \\
        Single-Agent w/ Hard Org. Spec. & 48.5 & 77.5 \\
        Multi-Agent w/o Org. Spec. & 43.5 & 82.0 \\
        Multi-Agent w/ Soft Org. Spec. & 38.8 & 86.0 \\
        \textbf{Multi-Agent w/ Hard Org. Spec. (KARMA)} & \textbf{33.0} & \textbf{90.7} \\
        \hline
    \end{tabular}}
\end{table}
\autoref{tab:adversarial_conditions} compares recovery times and service availability in the \textit{DDoS Attack} scenario. KARMA achieves the fastest recovery time (\textbf{33.0s}), outperforming AWARE (\textbf{38.8s}) and Rlad-core (\textbf{43.5s}). It also ensures a good service availability at \textbf{90.7\%}, reducing downtime compared to AWARE (\textbf{83.6\%}).

Traditional autoscalers like KHPA and Gym-HPA rely on reactive threshold-based scaling, leading to slower recovery and lower service availability under attacks. RL-based methods such as Rlad-core and AWARE improve resilience but lack structured coordination, making them less effective against adversarial spikes. Single-agent approaches struggle with balancing attack mitigation and resource optimization, while multi-agent models with soft constraints allow exploratory actions that sometimes delay optimal responses.

KARMA's proactive adversarial learning and structured coordination enable it to anticipate attacks rather than react after degradation. Explicit role-based constraints ensure agents prioritize critical scaling actions, resulting in faster mitigation and higher availability. These results highlight the effectiveness of multi-agent structured learning in security-sensitive autoscaling, where traditional methods exhibit slower adaptation and prolonged downtime.

\subsection{Gap 3: Digital Twin Modeling}

\begin{table}[h]
    \centering
    \caption{Transition models accuracy across all scenarios.}
    \label{tab:digital_twin_accuracy}{
        \footnotesize
    \begin{tabular}{>{\raggedright\arraybackslash}m{6cm}>{\centering\arraybackslash}m{2cm}}
        \hline
        \textbf{Baseline} & \textbf{Accuracy (\%)} \\
        \hline
        Without MLP Transition Model & 83.5 \\
        \textbf{With MLP Transition Model (KARMA)} & \textbf{94.9} \\
        \hline
    \end{tabular}}
\end{table}
The accuracy of the digital twin model is critical for training agents under realistic conditions.
\autoref{tab:digital_twin_accuracy} compares the accuracy of different digital twin models. The results show that KARMA achieves \textbf{94.9\%} accuracy, outperforming the non-MLP model (\textbf{83.5\%}), which struggles to generalize.

The improvement stems from the MLP model's ability to capture non-linear dependencies between workload fluctuations, resource allocation, and scaling actions. Without this feature, the system fails to model complex cluster behaviors accurately.
By leveraging a neural network for transition modeling, KARMA ensures a more reliable digital twin, allowing agents to train under near-realistic conditions. This reduces the risk of poor decision-making when policies transfer to production, reinforcing the importance of high-fidelity simulations.

\subsection{Gap 4: Automated MAS Generation}

\begin{figure}[h!]
    \centering
    \includegraphics[width=0.49\textwidth]{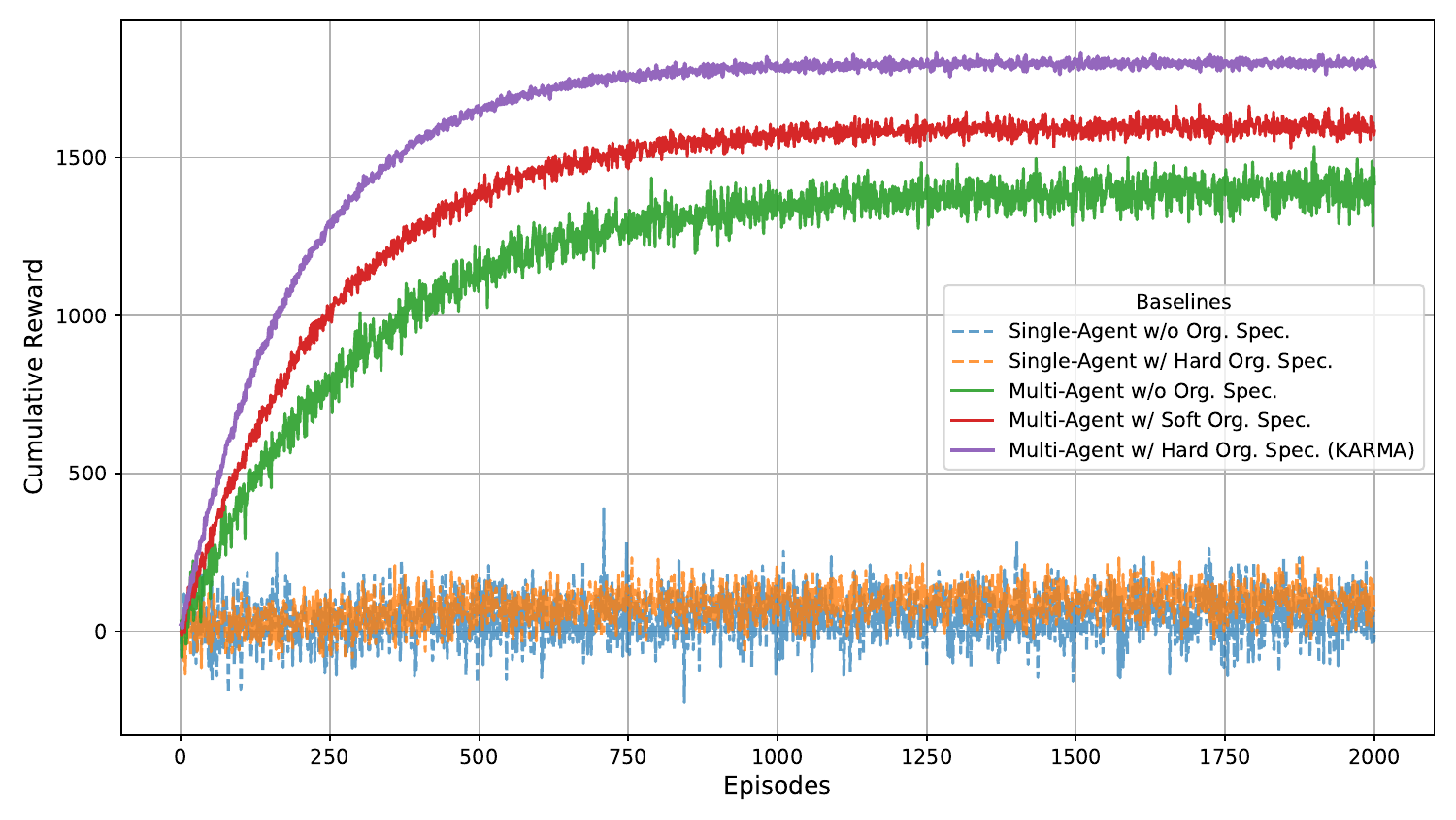}
    \caption{Learning curves across baselines for the mixed scenario over 2000 episodes.}
    \label{fig:learning_curves}
\end{figure}

The efficiency of generating a MAS is evaluated in terms of convergence time and training overhead. \autoref{tab:mas_generation_efficiency} presents the results across concerned scenarios while \autoref{fig:learning_curves} shows learning curves in the mixed scenario over 2000 episodes.

\begin{table}[h]
    \centering
    \caption{MAS generation efficiency across all scenarios.}
    \label{tab:mas_generation_efficiency}{
        \footnotesize
    \begin{tabular}{>{\raggedright\arraybackslash}m{3.5cm}>{\centering\arraybackslash}m{2cm}>{\centering\arraybackslash}m{2cm}}
        \hline
        \textbf{Baseline} & \textbf{Convergence Time (episodes)} & \textbf{Training Overhead (hours)} \\
        \hline
        Multi-Agent w/o Org. Spec. & 1800 & 4 \\
        \textbf{Multi-Agent w/ Hard Org. Spec. (KARMA)} & \textbf{950} & \textbf{1.5} \\
        \hline
    \end{tabular}}
\end{table}

The role-guided learning narrows the search space for optimal policies, enabling faster convergence and reduced computational costs. These efficiency gains are particularly important for scaling MAS solutions to complex environments.

The learning curves in \autoref{fig:learning_curves} demonstrate that KARMA achieves stable convergence significantly faster than the baseline without organizational specifications. By episode 950, KARMA exhibits minimal variance in cumulative rewards, whereas the baseline requires nearly double the episodes (1800) to reach comparable performance. This highlights the role of organizational constraints in guiding agents toward effective policies, thereby reducing exploration overhead.

\noindent \textit{Overhead Discussion:} The training phase for KARMA required approximately 1.5 hours on a high-performance machine with 1 GPU (Tesla V100, 16GB) and 16 CPU cores, converging in under 1,000 episodes. At inference time, each agent's policy produces decisions in under 30ms, with a negligible memory footprint (<50MB per agent).

\autoref{tab:mas_generation_efficiency}, shows reduced \textit{Convergence Time} by approximately \textbf{47\%} compared to Multi-Agent w/o Org. Spec., showcasing the efficiency of role-guided learning in minimizing unnecessary exploration. Moreover, \textit{Training Overhead} is reduced by \textbf{62.5\%}, showing a better practicality for large-scale systems where computational resources are a limiting factor.

\subsection{Gap 5: Adaptability}

Adaptability assesses the ability of the system to maintain performance under dynamic workloads accross scenarios.
\begin{table}[h]
    \centering
    \caption{Comparison of adaptability in the mixed scenario.}
    \label{tab:adaptability_comparison}{
    \footnotesize
    \begin{tabular}{>{\raggedright\arraybackslash}m{5cm}>{\centering\arraybackslash}m{3cm}}
        \hline
        \textbf{Baseline} & \textbf{Reward s.t.d (\%)} \\
        \hline
        Single-Agent w/o Org. Spec. & 11.1 \\
        Single-Agent w/ Hard Org. Spec. & 11.1 \\
        Multi-Agent w/o Org. Spec. & 10.7 \\
        Multi-Agent w/ Soft Org. Spec. & 9.0 \\
        \textbf{Multi-Agent w/ Hard Org. Spec. (KARMA)} & \textbf{5.3} \\
        \hline
    \end{tabular}}
\end{table}
\autoref{tab:adaptability_comparison} shows that KARMA achieves the lowest reward standard deviation (\textbf{5.3\%}), indicating a highly stable performance, outperforming Multi-Agent w/ Soft Org. Spec. (\textbf{9.0\%}) and Multi-Agent w/o Org. Spec. (\textbf{10.7\%}).

In the \textit{Mixed Scenario}, single-agent models exhibit higher variance as they must balance multiple competing objectives without specialization. Multi-agent approaches improve adaptability, but without structured coordination, fluctuations remain. The use of soft organizational constraints stabilizes performance, though some exploratory variations persist.

KARMA's hierarchical reinforcement learning ensures lower variability by decomposing the overarching goal into specialized sub-goals. This structured approach enables agents to focus on well-defined objectives, reducing conflicting decisions and improving overall stability.
These findings underscore the importance of structured learning frameworks in autoscaling. By enforcing clear agent specializations, KARMA enhances adaptability, ensuring resilient performance across unpredictable workload conditions.

\subsection{Gap 6: Explainability}
\label{subsec:gap_explainability}

\begin{figure}[h!]
    \centering
    \includegraphics[width=0.49\textwidth]{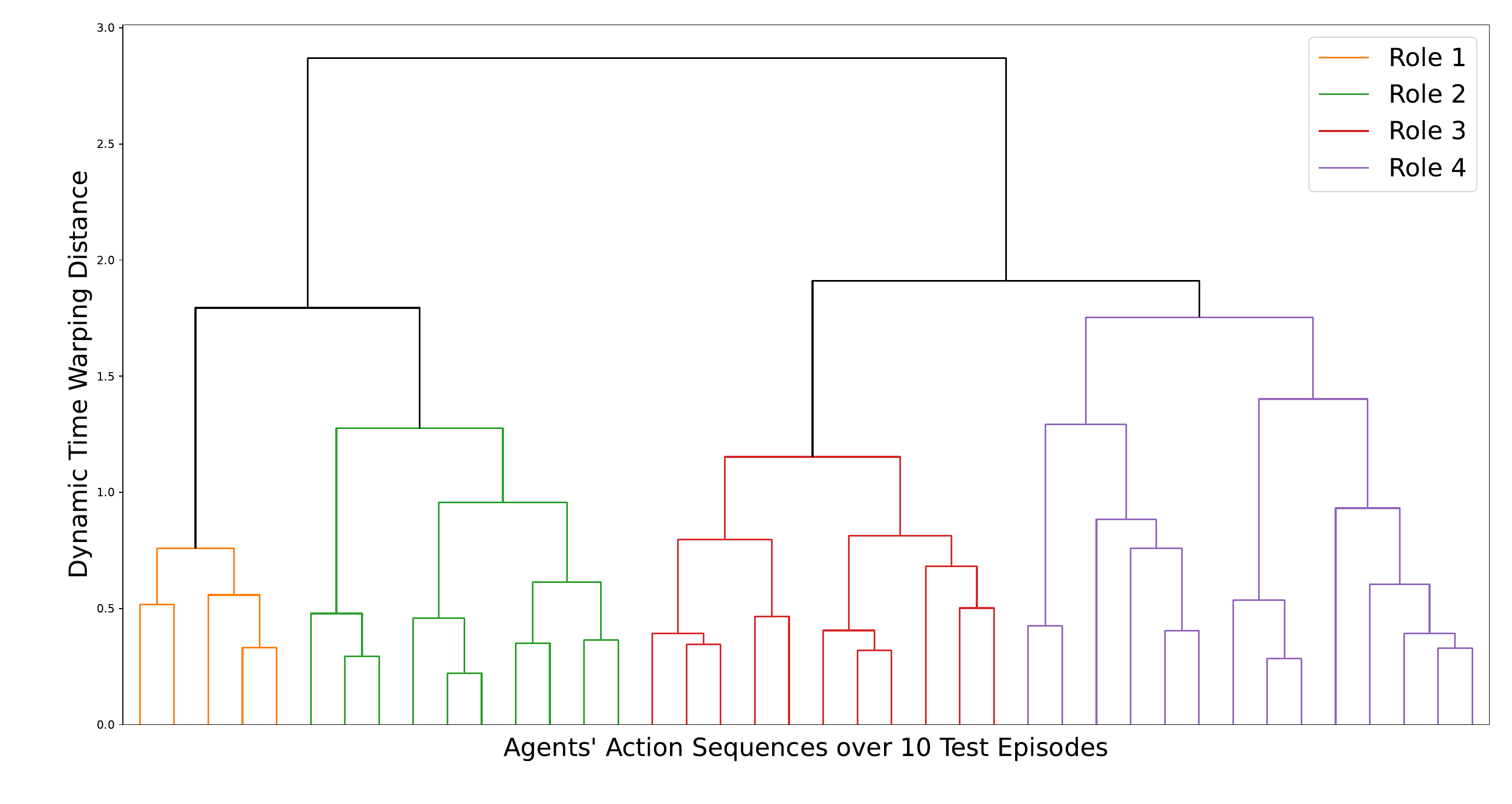}
    \caption{Dendrogram obtained after hierarchical clustering of agent trajectories for role inference in the mixed scenario.}
    \label{fig:trajectory_clustering_hrl}
\end{figure}

Explainability is qualitatively evaluated through trajectory clustering and quantitatively through the alignment of agent behaviors with predefined roles and missions.
\noindent \autoref{fig:trajectory_clustering_hrl} illustrates the dendrogram generated by hierarchical clustering of agents' action sequences with the four roles applied, using DTW as the similarity measure. The figure highlights the emergence of four distinct clusters, each corresponding to a specific organizational role, demonstrating the ability of the agents' behaviors to align with the predefined roles.

\begin{table}[h!]
    \centering
    \caption{Alignment of agents with roles and missions.}
    \label{tab:alignment}
    {\footnotesize
    \begin{tabular}{>{\raggedright\arraybackslash}m{4.1cm}>{\centering\arraybackslash}m{1.5cm}>
    {\centering\arraybackslash}m{1.5cm}}
    \toprule
    \textbf{Baseline} & \textbf{Alignment Score (\%)} & \textbf{Clustering Purity (\%)} \\
    \midrule
    Multi-Agent w/o Org. Spec. & $\emptyset$ & 62.7 \\
    Multi-Agent w/ Soft Org. Spec. & 85.3 & 70.1 \\
    \textbf{Multi-Agent w/ Hard Org. Spec. (KARMA)} & \textbf{96.2} & \textbf{89.4} \\
    \bottomrule
    \end{tabular}
    }
\end{table}

KARMA shows the emergence of distinct behavioral patterns aligned with predefined roles validates KARMA's organizational model and has the highest alignment score (\textbf{96.2\%}), significantly outperforming Multi-Agent w/o Org. Spec. (\textbf{85.3\%}), showcasing well-coordinated agent behaviors. In adversarial scenarios, clustering purity is highest for KARMA, reflecting the clear differentiation of agent behaviors under organizational constraints.
Distinct clusters validate the role-specific behaviors and highlight the interpretability of agents. Baselines without organizational specifications show reduced explainability, as evidenced by lower clustering purity and alignment scores with soft organizational constraints.




\section{Conclusion}
\label{sec:conclusion}

This paper presented KARMA, a framework aimed at improving the operational resilience of Kubernetes clusters. While modular designs offer simplicity, they often rely on manual coordination that struggles in dynamic or adversarial contexts. In contrast, KARMA’s MAS approach enables adaptive, decentralized responses through agent specialization.
The experimental results demonstrate that KARMA effectively addresses several critical gaps in Kubernetes autoscaling. By integrating MARL with organizational principles, KARMA achieves improvements in adversarial robustness, and explainability. Its ability to decompose complex goals into roles and missions ensures coordinated agent behavior. The use of MLP-based transition model, further strengthens KARMA's capacity to simulate realistic conditions.
%
%

However, some aspects need to be further explored:
\begin{enumerate*}[label=\textbf{\arabic*)}, itemjoin={;\quad }]
    \item \textbf{Simulation-to-Reality Gap:} Even though we generate a near-realistic simulation model from environment traces, we need to better simulate unaccounted unexpected system failures
    \item \textbf{Dependence on Domain Expertise:} Defining roles, missions, and reward structures relies heavily on domain-specific knowledge, which may limit the framework's generalizability
    \item \textbf{Computational Overhead:} The training process with multi-agent configurations and organizational constraints requires substantial computational resources.
    \item \textbf{Scalability to Multi-Node Clusters:} Although current experiments focus on a single-node cluster, preliminary evaluations are being conducted on larger-scale deployments.

\end{enumerate*}


\section*{Acknowledgment}
    \textit{GPT-4o} was only used for spelling corrections.

\section*{References}

\bibliographystyle{IEEEtran}
\bibliography{references}

\end{document}